\documentclass[12pt]{iopart}

\usepackage{iopams}
\usepackage{graphicx}
\usepackage[english]{babel}
\usepackage{graphics}
\usepackage{rotating}
\usepackage[usenames]{color}
\usepackage{ulem}

\newcommand{\dnd}{$\langle (\Delta N)^2 \rangle$}
\newcommand{\dndp}{$\langle (\Delta N)^2 \rangle_{\rm p}$}
\newcommand{\dndn}{$\langle (\Delta N)^2 \rangle_{\rm n}$}

\begin{document}
\title[Self-consistent HF+BCS calculations with finite-range interactions]
{A study of self-consistent Hartree--Fock plus Bardeen--Cooper--Schrieffer 
calculations with finite-range interactions}.
  
\author{M. Anguiano, A. M. Lallena}
\address{Departamento de F\'{\i}sica At\'omica, Molecular y
  Nuclear, Universidad de Granada, E-18071 Granada, Spain}
\ead{mangui@ugr.es}
\author{G. Co', V. De Donno}
\address{Dipartimento di Matematica e Fisica ``E. De Giorgi'',
 Universit\`a del Salento and,
 INFN Sezione di Lecce, Via Arnesano, I-73100 Lecce, Italy}


\begin{abstract}
In this work we test the validity of a Hartree--Fock plus Bardeen--Cooper--Schrieffer
  model in which a finite-range interaction
  is used in the two steps of the calculation by comparing the results obtained to those found in
  a fully self--consistent Hartree--Fock--Bogoliubov calculations using the same interaction. 
  Specifically, we consider the Gogny--type D1S and D1M forces. We study a wide range of spherical
  nuclei, far from the stability line, in various regions of the nuclear chart, from
  oxygen to tin isotopes. We calculate various
  quantities related to the ground state properties of these nuclei, such as
  binding energies, radii, charge and density distributions and elastic
  electron scattering cross sections. 
  The pairing effects are studied by direct comparison with the Hartree-Fock results.
Despite of its relative simplicity, in most of the cases, our model provides results very close 
to those of the  Hartree--Fock--Bogoliubov calculations, and it reproduces rather well the empirical 
evidences of pairing effects in the nuclei investigated.
\end{abstract}

\vskip 0.5cm 

\maketitle
%

\section{Introduction}
\label{sec:int}

The study of the structure of nuclei in regions of the nuclear chart
which have not yet been experimentally explored can identify new
nuclear properties, and deviations from expected behaviours based on
previously established rules~\cite{sat03}. Even though great
attention has been addressed to the study of the excitation spectrum
of these nuclei~\cite{dob07,urb00,bla05}, one should not neglect the
observables related to their ground states such as binding energies,
proton and neutron densities, and single particle (s.p.) energies.

In nuclei far from the shell closure, the role of the pairing
correlations is particularly important.  The simplest mean-field
approach taking into account these correlations is the
Bardeen--Cooper--Schrieffer (BCS) model
\cite{boh58,row70,rin80,suh07}. This approach is based on a set of
s.p. states given as input. By using a pairing
nucleon-nucleon interaction, the model generates partial occupation
probabilities for each s.p. state. These probabilites are used to
evaluate the pairing effects on the various ground state observables.

The first implementations of the BCS model used s.p. wave
functions obtained with phenomenological potentials
\cite{bar60,kis63,row70}.  More recently, also wave
functions generated by Hartree-Fock (HF) calculations have been considered 
\cite{rob87,bon05,sar08,ber09,ber12}. This latter approach is 
called HF+BCS method.
While in this approach the production of the s.p. states
and the evaluation of their pairing correlations are treated in two different types of 
  calculations, the two processes are unified in the
  Hartree--Fock--Bogoliubov (HFB) theory   \cite{rin80}.

The two approaches quoted above,
HF+BCS and HFB, are mean-field models that make use of effective forces,
i.e. forces whose parameters have been chosen to reproduce nuclear properties
within mean-field theories. Among these effective forces,  the zero-range Skyrme interaction, 
in various parameterizations, has been widely used in both 
HF+BCS \cite{rob87,bon05,sar08,ber09,ber12} and HFB
\cite{ter96,gra01,ben05} calculations. 
These calculations are not self-consistent since 
the zero-range character of the Skyrme interaction generates results
strongly dependent on the size of the s.p. configuration space, therefore a different zero-range interaction must 
be used in the pairing channel. 
  The parameters of this zero-range pairing
  interaction are chosen to reproduce some specific pairing
  properties within the  particular s.p. configuration space used
  in the calculation. 

As widely discussed in the seminal paper of Decharg\'e and Gogny
\cite{dec80}, this is not any more necessary when
finite-range interactions are used, because they provide an automatic
convergence of the results, once a sufficiently large  size of the
s.p. configuration space is adopted, and  this allows really
self-consistent calculations.

There are many examples of HFB calculations 
carried out with finite-range interactions. In these works the
D1 \cite{dec80}, D1S \cite{ber91,ang01a,nak06,rob12}, 
D1M \cite{gor09} Gogny interactions and the M3Y force
\cite{nak08} have been used. On the other hand, we know only few works where
finite-range interactions are used in HF+BCS calculations.  This
approach has been used by Robledo {\it et al.} \cite{rob87} with the
D1 interaction to analyze the octupole excitation in actinide
isotopes, and more recently by Than {\it et al.} \cite{tha11} with the
D1S and the M3Y interactions to study the inner crust of neutron
stars.
The purpose of the present article is to explore, and exploit, the features offered
by a HF+BCS approach with finite-range interactions. The use of an
unique interaction for the HF and pairing calculations offers a large
prediction power, and this is associated with
the advantage of calculations which are numerically less involved than
those of the HFB approach. One of the goals of this article is to test the limits of this self--consistent HF+BCS
approximation for studying pairing correlations in regions of
the nuclear chart far from the stability line. For this purpose, we compare the results of both
HF+BCS and HFB approaches by using the same finite-range effective nucleon-nucleon interaction.

In section \ref{sec:model}, we briefly present the basic ideas of our
approach, and we describe how we calculate the various observables we
have considered in our work. In section \ref{sec:result} we present our
results. First we compare the fluctuation of the particle number
obtained in our model 
with that of the HFB to test the validity of
our calculations. Then, we discuss the results obtained for binding
energies, s.p. properties, proton and
neutron root mean square (rms) radii and density distributions. In
section \ref{sec:conclusion} we summarize our results and we draw our
conclusions. 

\section{The model}
\label{sec:model}
In this section we present the main features of our approach, and
describe how we calculate binding energies, radii, and density
distributions. The first step of our
HF+BCS procedure consists in obtaining the s.p. wave functions $\phi_i$ by
solving the HF equations \cite{rin80} 
\begin{equation}
-\frac{\hbar^2}{2m} \, \nabla^2  \phi_i({\bf r}_1) \,
+ \, U({\bf r}_1) \, \phi_i({\bf r}_1) \, -\, \int {\rm d}^3r_2 \, W({\bf r}_1,{\bf r}_2) \, \phi_i({\bf r}_2) 
\, = \, \epsilon_i  \, \phi_i({\bf r}_1) \, ,
\label{eq:HF}
\end{equation}
where $m$ is the nucleon mass and $\epsilon_i$ the s.p. energy. In the
above equation we have indicated the Hartree term as
\begin{equation}
 U({\bf r}_1) \, =\,  \int {\rm d}^3r_2 \, V({\bf r}_1,{\bf r}_2) \, 
\sum_{k \le \epsilon_{F}} |\phi_{k}({\bf r}_2)|^2
\, ,
\label{eq:hartree}
\end{equation}
and the Fock--Dirac term as
\begin{equation}
W({\bf r}_1,{\bf r}_2) \,= \, \sum_{k \le \epsilon_F} \,V({\bf r}_1,{\bf r}_2) \, 
\phi^*_{k}({\bf r}_2) \, \phi_{k}({\bf r}_1) 
\, .
\label{eq:fock}
\end{equation}
Here $\epsilon_F$ is the Fermi energy, and $V({\bf r}_1,{\bf r}_2)$ the
effective nucleon-nucleon interaction.

In this work, we consider only spherical even-even nuclei  and 
solve the HF equations with the method described in Refs. \cite{co98b,bau99}. Once the iteration
procedure has reached convergence, the potential terms
(\ref{eq:hartree}) and (\ref{eq:fock}) are used to generate a set of
s.p. wave functions which includes all the states below the Fermi level
and a large set of states above it. Each s.p. state $|k\rangle \equiv
|\epsilon_k ;n_k l_k j_k m_k \rangle$ is characterized by the
principal quantum number $n_k$, the orbital angular
momentum $l_k$, the total angular momentum $j_k$, its $z$ axis
projection $m_k$ and, finally, the $2j_k+1$ degenerated energy
$\epsilon_k$. These s.p. wave functions and energies are used to solve the BCS
equations.

In the BCS theory the ground state of the system is defined as
\begin{equation}
|{\rm BCS}  \rangle \, = \, \prod_{k >0 } \left ( u_{k} \,+ \, v_{k} \, a_{k}^{\dagger} 
\, a_{\bar{k}}^{\dagger} \right ) \, |-\rangle \, ,
\label{eq:BCS}
\end{equation}
where $| - \rangle$ is the vacuum, 
$a_{k}^{\dagger}$ is the creation operator of the state $|k\rangle$, $a_{\bar{k}}^{\dagger}$ 
is the creation operator of the state 
\begin{equation}
|\bar{k}\rangle \,= \, (-1)^{j_k-m_k} \, |-k\rangle \, = \, (-1)^{j_k-m_k} \, |\epsilon_k ;n_k l_k j_k -m_k\rangle  \, ,
\label{eq:kbar}
\end{equation}
and $|v_{k}|^2$ the
probability that the state $|k\rangle$ is occupied, and is related to
$u_k$ by the condition
\begin{equation}
|u_{k}|^2 +  |v_{k}|^2 = 1 \, .
\end{equation}
It is worth to point out that $u_{\bar{k}}=u_k$ and $v_{\bar{k}}=-v_k$.

The BCS state defined in 
equation (\ref{eq:BCS}) is not an eigenstate of the particle
number operator. For this reason, the BCS equations are obtained by applying the variational principle to
the expectation value of the operator
\begin{equation}
\mathcal{H} \, = \, H \, - \, \lambda N \, ,
\end{equation}
where $H$ is the traditional nuclear hamiltonian, $N$ is the particle
number operator and $\lambda$ is a Lagrange multiplier.
The parameters to be changed in the variational procedure are the
$v_k$ and $u_k$ coefficients. The application of the variational principle leads to solve the set of equations
\begin{equation}
\left ( u_k^2 -v_k^2 \right ) \,  \Delta_k \, = \, 2\, u_k\, v_k\, \eta_k \, ,
\label{eq:bcs1}
\end{equation}
where 
\begin{equation}
\Delta_k \, = \,  - \frac{1}{{\widehat{j}}_k} \sum_i {\widehat{j}}_i\, u_i\, v_i \,  
\langle ii \, 0| {V}| kk\,  0 \rangle  
\, , 
\label{eq:bcs2}
\end{equation}
and
\begin{equation}
\eta_k \, = \, \langle k|T|k \rangle \,- \, \lambda \, .
\label{eq:bcs3}
\end{equation}
In the last definition, we have indicated with
$T$ the kinetic energy operator and we dropped a
renormalization term \cite{suh07}. In the above equations, for the indexes indicating
angular momentum eigenvalues, we have used the symbol 
$\widehat{j} =\sqrt{2 j+1}$.   
The matrix element of the interaction in equation (\ref{eq:bcs2}) indicates that the s.p. wave 
functions are coupled to global angular momentum zero.

The solution of the BCS equations, Eqs.~(\ref{eq:bcs1})-(\ref{eq:bcs3}),
provides the values of $v_k$ and $u_k$ for each s.p. state
included in the configuration space.  The knowledge
of $v_k$ and $u_k$ and $\phi_k$ allows the evaluation
of the expectation values of various ground state quantities related
to the BCS ground state. 

We have considered the particle number fluctuation index which is defined as 
\begin{equation}
\langle (\Delta N)^2 \rangle \,  = \, \langle {\rm BCS} | \hat{N}^2| {\rm BCS } \rangle \,-\, \langle {\rm BCS } |
\hat{N} | {\rm BCS} \rangle^2 \, =\,  4 \sum_{k} (2j_k+1) \, u_{k} \, v_{k}   \, .
\label{eq:dn2}
\end{equation}
The value of this quantity is directly related to the relevance of the
pairing effects. 
In what follows, we indicate as
$\langle (\Delta N)^2 \rangle_{\rm p}$ and 
$\langle (\Delta N)^2 \rangle_{\rm n}$ the values obtained when
the sum of equation (\ref{eq:dn2}) is restricted to
proton or neutron s.p. states, respectively.
In the next section we shall compare the values
obtained in our HF+BCS approach with those of the HFB calculation. 

We calculate the quasi-particle energy as
\begin{equation}
E_k \equiv \sqrt {\epsilon_k^2 + \Delta_k^2} \, ,
\label{eq:qpenergy}
\end{equation}
and the global energy of the HF+BCS ground state as
\begin{equation}
E \,=\,  \frac{1}{2}\sum_k (2j_k+1) \, 
\left[ v_k^2 (\epsilon_k +\langle k|T|k \rangle) \,-\, u_k \,v_k\, \Delta_k \right] \, .
\end{equation}

We evaluate also the HF+BCS matter densities as
\begin{equation}
\rho(r) \,=\, \sum_{k} v_{k}^2 |\phi_{k} ({\bf r}) |^2 \, =\, \frac{1}{4 \pi}
\sum_k (2j_k+1) \, v_k^2 |R_k(r) | ^2 
\, ,
\label{eq:rho}
\end{equation}
where $R_k$ denotes the radial part of the s.p. wave function normalized as
\begin{equation}
\int_0^\infty \,r^2 |R_k(r) | ^2 \, dr =1
\,\,.
\end{equation}
With the proper limitations in the sum of 
equation (\ref{eq:rho}) we calculate the proton
and neutron density distributions. The charge distribution is obtained by folding the
proton distribution with the nucleon electromagnetic form factor. By
using the matter distributions we determine the rms
radii as 
\begin{equation}
R_\alpha \,=\, \langle r^2 \rangle^{1/2}_\alpha \,=\, 
\left[
\frac{\displaystyle \int_0^\infty r^4 \, \rho_\alpha(r) \, {\rm d} r }
     {\displaystyle \int_0^\infty r^2 \rho_\alpha(r) \, {\rm d} r } \right]^{1/2} \, ,
\end{equation}
where $\alpha$=p
indicates protons, and $\alpha$=n neutrons. 

The HF+BCS densities differ from the HF ones because of the changes in the 
occupation probabilities induced by the BCS calculation. This is a remarkable 
difference with respect to HFB where both s. p. wave functions and their 
occupation probabilities are obtained in a self-consistent way.

As already pointed out in the introduction our HF+BCS calculations
have been carried out by using an effective finite-range interaction of
Gogny type  \cite{dec80,ber91}, which can be expressed 
as a sum of a central, $V_{\rm C}$, a spin-orbit, $V_{\rm SO}$, and a density
dependent, $V_{\rm  DD}$, term
\begin{equation}
V(1,2)\, = \, V_{\rm C}(1,2)\,+\,V_{\rm SO}(1,2)\,+\,V_{\rm DD}(1,2)\, .
\label{eq:vgog}
\end{equation}
Only the central term has a finite range.  
The other two terms are of zero-range type, and have the same expressions
  used in Skyrme interactions. While in the HF calculations the full
  expression of the interaction (\ref{eq:vgog}) is used, in the BCS calculations we
  consider only the central, finite-range term. 
This is the procedure commonly adopted in the pairing sector of 
the HFB calculations  when 
   Gogny type interactions are used \cite{dec80,ber91,egi95}. 

Our calculations have been carried out with
two parametrizations of the Gogny 
force. A first one is the more traditional D1S 
\cite{ber91}, well known and widely used in the literature. The second
one is the more recent D1M \cite{gor09} built as a correction of the D1S force
to have a realistic behaviour of the neutron matter 
equation of state above saturation densities.

We have already remarked that the BCS calculations are carried out
after the s.p. configuration space has been chosen. The size of this
space should be large enough to ensure the stability of the
results. This implies that some of the s.p. states, obviously above the
Fermi energy, lie in the continuum. To the best of our knowledge,
there is not a BCS approach treating the continuum part of the
configuration space without approximation, contrary to what has been
done with the Random Phase Approximation \cite{don08t}. We have adopted a procedure where the continuum
part of the s.p. configuration space is conveniently discretised
\cite{co98b,bau99,ang11}. Specifically, we have chosen a box of radius 
$R_{\rm max}=4\,r_0\,A^{1/3}$, with $r_0=1.2$ fm, where the energy levels 
were calculated with the HF potential.

From the numerical point of view, the sum on the s.p. states in equation (\ref{eq:bcs2}) is
automatically limited when a finite-range interaction is used. 
In our BCS calculations, we have included all states with energy below 10 MeV for all
nuclei studied. We have checked that this guarantees the convergence 
of the BCS energy with an order of accuracy in the keV range.

\section{Results}
\label{sec:result}
We have applied
our HF+BCS approach to study
oxygen, calcium, nickel, zirconium and tin isotope chains. 
We have also investigated the $N=40$ and $N=50$
isotone chains. All nuclei studied are of even--even type. 
In some cases we compare our results with those obtained 
in HFB calculations
  \cite{egi95,ang01a}. 

\subsection{Particle number fluctuation}

In this section we compare the pairing correlations described by our HF+BCS approach
with those obtained by the HFB model.
We conduct this investigation by considering the particle number fluctuation index
\dnd, defined in equation (\ref{eq:dn2}), a quantity particularly sensitive
to the pairing correlations. We compare the results obtained with our
HF+BCS model with those obtained in HFB calculations by using the same
interaction. 

We start our analysis by investigating the chains of nuclei where one
type of nucleons fills
completely all the s.p. levels below the Fermi surface. For these nucleons
the pairing effects 
are negligible, therefore
$\langle (\Delta N)^2 \rangle_{\alpha}$, $\alpha \equiv {\rm p}$ or n, 
is zero.

\begin{figure}
\centering
\includegraphics[scale=0.8]{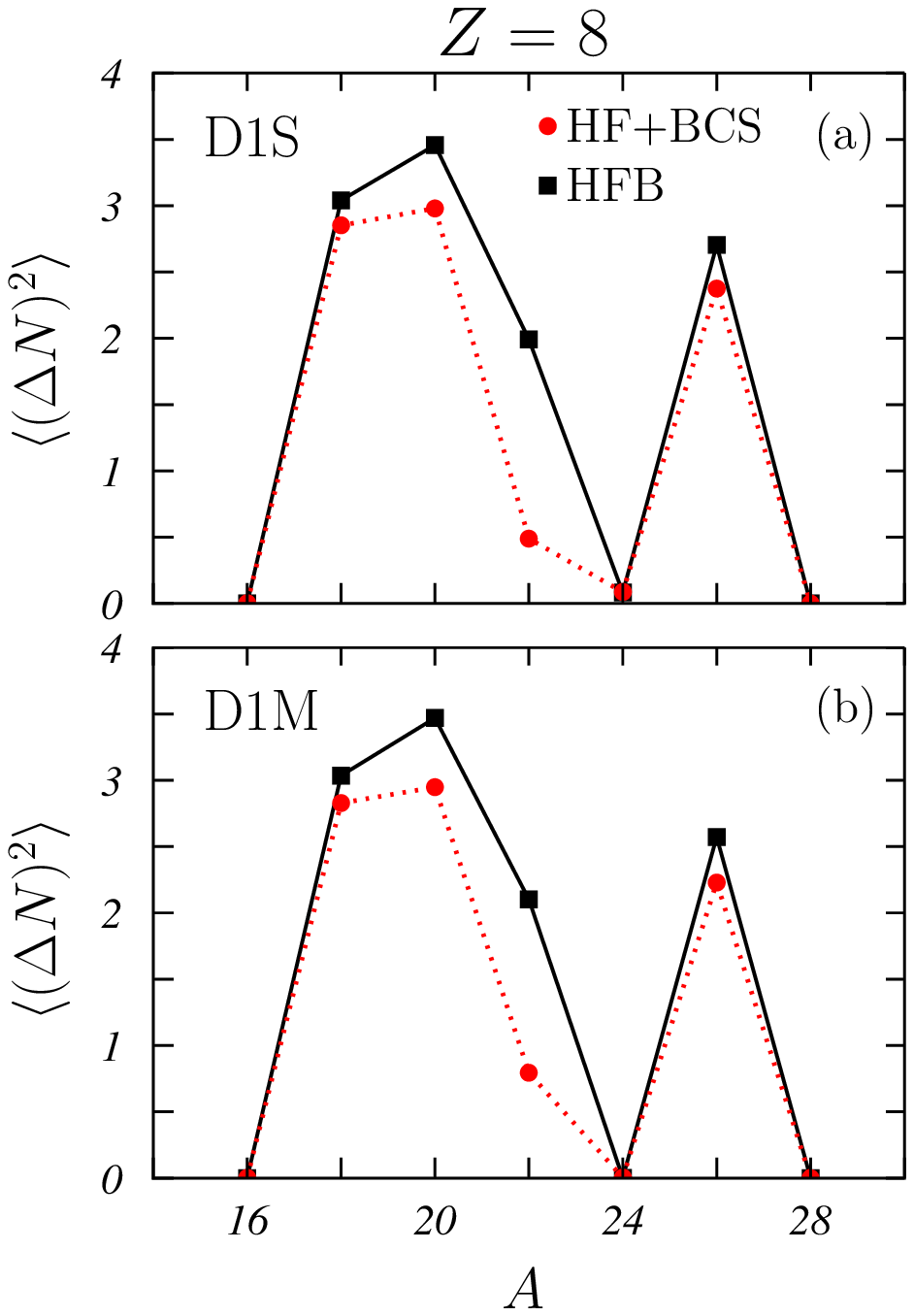} 
\caption{\small (Color on line) 
Values of the particle fluctuation index \dnd \/ in oxygen isotopes obtained in HF+BCS (red dots) and 
HFB (black squares) calculations carried out with the 
D1S, panel (a), and D1M, panel (b), Gogny interactions.
The lines are drawn to guide the eyes. In this case only neutrons contribute, therefore
 \dnd$\equiv$\dndn.
\label{fig:DN2-Z8}
}
\end{figure}

We show in figure \ref{fig:DN2-Z8} the \dnd \/  values, as a function of the mass number, for a set of
oxygen isotopes calculated with the D1S and D1M  interactions, panel (a) and (b), respectively. For these isotopes \dndp \/ is zero, since $Z=8$ closes
a shell. In the figure, the red dots show our HF+BCS results
while the black squares those obtained in
HFB calculations. The lines have been drawn to guide the eyes.
We observe that the two interactions produce very
similar results.  For this reason, in the remaining part of this
section we shall present only the values found with the D1M
interaction, unless explicitly stated.

\begin{figure}
\centering
\includegraphics[scale=0.8]{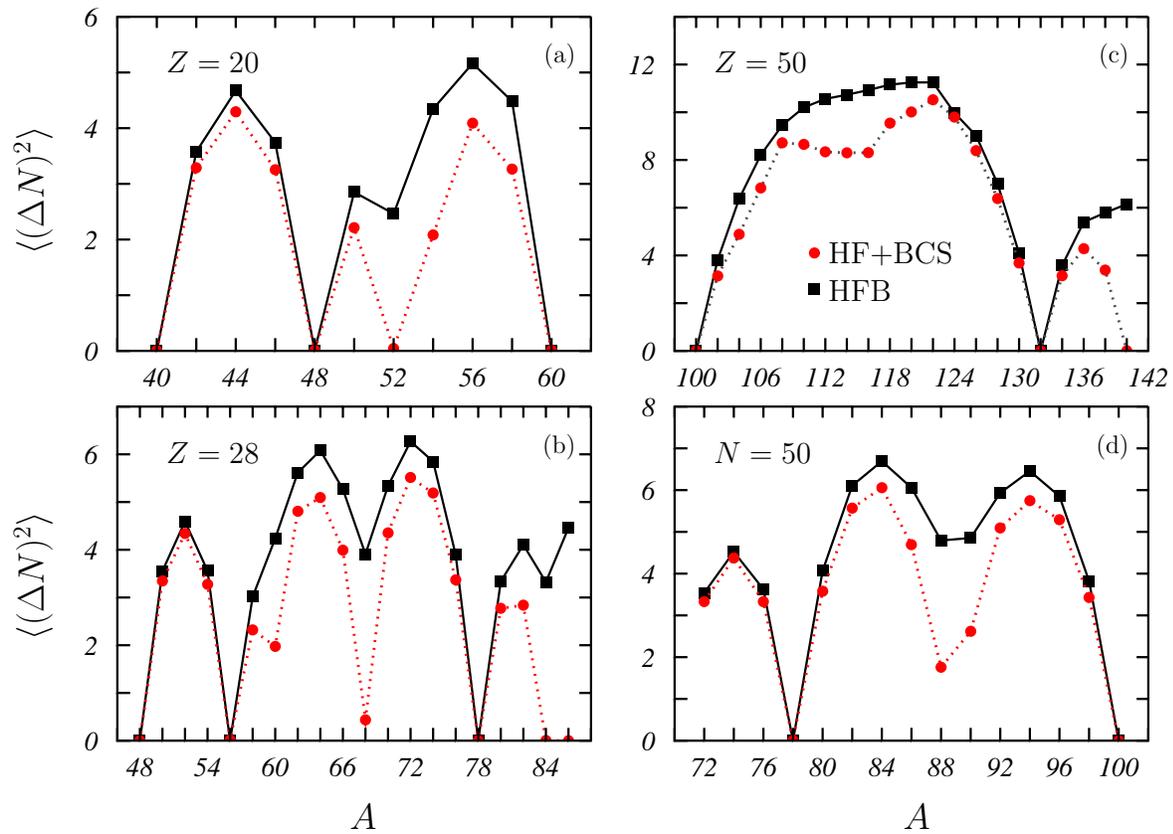} 
\caption{\small (Color on line) Values of 
\dnd \/  for calcium (a),  nickel (b), and tin (c) isotopes and 
 $N=50$ isotones (d) obtained in HF+BCS (red circles) and HFB 
 (black squares) calculations with the Gogny D1M interaction. 
 In panels (a)-(c) only neutrons contribute and \dnd$\equiv$\dndn. In panel (d) only protons 
 contribute and \dnd$\equiv$\dndp.
\label{fig:DN2}
}
\end{figure}

Together with the oxygen results of figure \ref{fig:DN2-Z8} we discuss
also the results shown in figure \ref{fig:DN2} for calcium, nickel and
tin isotopes and  a set of isotones with $N=50$. 
The behaviour of the results of the two types of calculations is very
similar in all the nuclear chains we have considered. 
However, all the \dnd \/ values obtained with our HF+BCS approach are smaller, or
at most equal, than those of the HFB calculations.  This is a clear
indication that, for a given interaction,  
the amount of pairing correlations generated by the HF+BCS approach
is smaller than that produced by the HFB calculations.

In the panel (a) of figure~\ref{fig:DN2}, where the results relative to the 
calcium isotopes are shown, the case of the $^{52}$Ca nucleus is remarkable. 
In this nucleus our HF+BCS  model predicts zero pairing, contrary to the result obtained in the 
HFB calculation.
In the HF description of the $^{52}$Ca nucleus,  
the neutron 2p$_{3/2}$ s.p. state at -5.56 MeV is fully occupied, and the 
next two s.p. levels are the 2p$_{1/2}$ state, whose energy is 2.3 MeV larger, and 
the 1f$_{5/2}$ state which has an even larger energy, 3.6 MeV. 
These energy differences are large enough to hinder, in BCS calculations, the mixing between these 
three s.p. levels.  On the contrary, the HFB calculations are more flexible and  generate pairing 
effects even between these so well separated s.p. states. 
As we are going to comment in the following, we have verified that the source 
of all the remarkable deviations between the HFB and our HF+BCS results
is due to situations rather similar to the one just discussed. 

Panel (b) shows the results obtained for the nickel isotopes, where the situation of the $^{60}$Ni nucleus is remarkable. 
This nucleus has the same 
neutron structure as the $^{52}$Ca nucleus but the \dndn \/ value obtained in the HF+BCS calculation shows 
a clear difference with respect to that of the HFB calculation. In this case, the pairing effects are not zero as in 
the  $^{52}$Ca case, even though they are noticeably smaller than those generated by the HFB calculation.

The discrepancy between the two calculations
is  evident for the $^{84}$Ni and $^{86}$Ni nuclei  
where all the HF s.p. states up to the
2d$_{5/2}$ and 3s$_{1/2}$ levels, respectively, are completely filled. 
The energy differences with respect to the next levels are about 1-2 MeV, and our BCS
calculations are unable to generate pairing correlations between these states. 

The results of panel (c) relative to the tin isotopes show differences between the two calculations
in a relatively large band of isotopes from $^{112}$Sn up to  $^{124}$Sn. 
These effects have consequences also on the global energies of these isotopes, and we shall 
discuss in detail their source in the next section. 
For larger values of the neutron number, we observe that
larges differences appear for the  $^{138}$Sn and $^{140}$Sn nuclei. In the last case neutrons
occupy all the s.p. states up to the 2f$_{7/2}$ level, but the closest s.p. states above it 
are situated at energies too large to allow the BCS calculations to generate enough pairing.

In the panel (d) of the figure we show the \dndp \/ values for a set of isotones with $N=50$.
The agreement between the results of the two approaches is quite good, with the exceptions of the 
$^{88}$Sr and $^{90}$Zr nuclei where the pairing effects in our HF+BCS model are smaller than those 
found in the HFB calculations. 
In the HF description of  these two nuclei,  
the proton levels are completely filled up to the  1f$_{5/2}$  state in 
$^{88}$Sr, and  up to the  2p$_{1/2}$ state in $^{90}$Zr. 
These results show, again, the difficulty of the HF+BCS approach to generate configuration
mixing analogous to that of the HFB model. 

\begin{figure}
\centering
\includegraphics[scale=0.8]{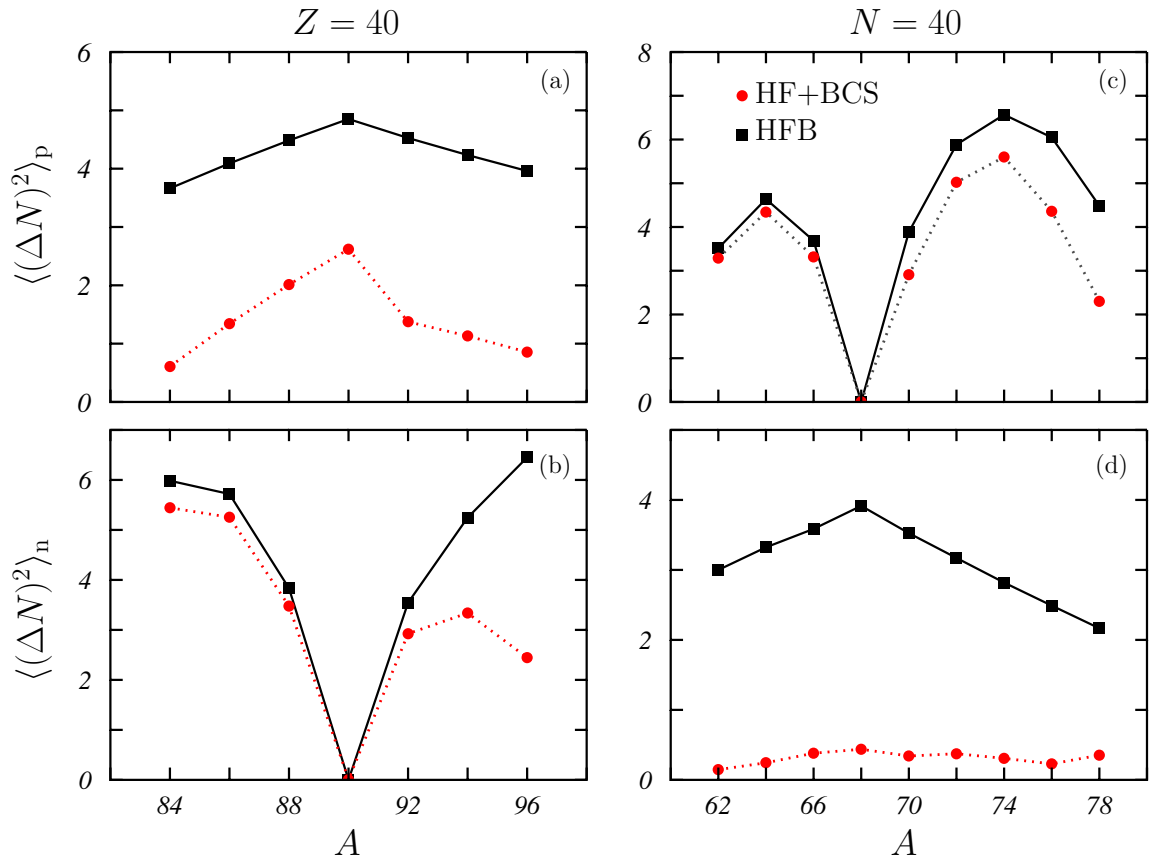} 
\caption{\small (Color on line) Values of
\dndp \/ (upper panels) and \dndn \/ (lower panels) 
for $Z=40$ isotopes (left panels) and $N=40$ isotones (rigth panels).
The meaning of the symbols is the same as in figure~\ref{fig:DN2}.
\label{fig:DN2-open}
}
\end{figure}

We show in figure \ref{fig:DN2-open}  the \dnd \/ values for nuclei where, in HF model, both
 proton and neutron levels are open. In the panels (a) and (b) we show, respectively, the \dndp \/
 and the \dndn \/ values for a set of  zirconium isotopes, $Z=40$, and in the other two panels we show the same
 quantities calculated for a set of $N=40$ isotones. Before entering in a detailed discussion of the 
 results, we want to remark that the HFB results are always larger, or at most equal, to those of
 the HF+BCS calculation.  
 
In a HF picture of zirconium isotopes, the protons fill completely all the s.p. levels up to the 2p$_{1/2}$ state.
The pairing interaction generates a mixing of the
1f$_{5/2}$, 2p$_{3/2}$, 2p$_{1/2}$ and 1g$_{9/2}$ proton s.p. levels
as the results shown in the panel (a) indicate.
The behaviour of the \dndp \/ values is rather flat, as we expected because
the proton number does not change. The HF+BCS and HFB results show similar behaviour
but rather different size. 
We find an analogous situation in the \dndn \/ results obtained for the $N=40$ isotones, as it
is shown in the panel (d) of the same figure. Also in this case the behaviour of the \dndn \/ values is rather
flat, and the values of the HF+BCS results are much smaller than those obtained in HFB calculations. 
The results of panel (a) and (d) indicate that the \dnd \/ values of the HF+BCS calculations are 
noticeably smaller than those obtained in HFB calculations when the number of one type of nucleons is 40. 
We point out that these are the largest
differences between HF+BCS and HFB results we have encountered in
our investigation. 

The situation is completely different for the other type of nucleons, as it is shown by the 
results presented in the panels (b) and (c) of the figure. 
In both cases the behaviour is not flat, furthermore, the differences between the \dnd \/ generated
by the two types of calculation are not as large as in the previously discussed cases. 

In panel (b) we show the \dndn \/ values for the zirconium isotopes. We observe that both calculations recognize
the shell closure for  $N=50$, corresponding to the $^{90}$Zr isotope. 
The situation is analogous for the results of panel (c) related to the $N=40$ 
isotones where the closure for the 28 protons, corresponding to the $^{68}$Ni nucleus, 
is well recognized  by the two  types of calculations. 

\subsection{Observables}

In this section we present the results of our study of the pairing effects on binding energies, 
and charge and matter density
distributions. We use the latter quantities to calculate rms radii
and elastic electron scattering cross sections. Our investigation has
been conducted by using both D1S and D1M interactions. Since the two
interactions provide similar results, to simplify the discussion, we present only those obtained with the 
D1M interaction.

\subsubsection{Binding energies.}

\begin{figure}
\centering
\includegraphics[scale=0.8]{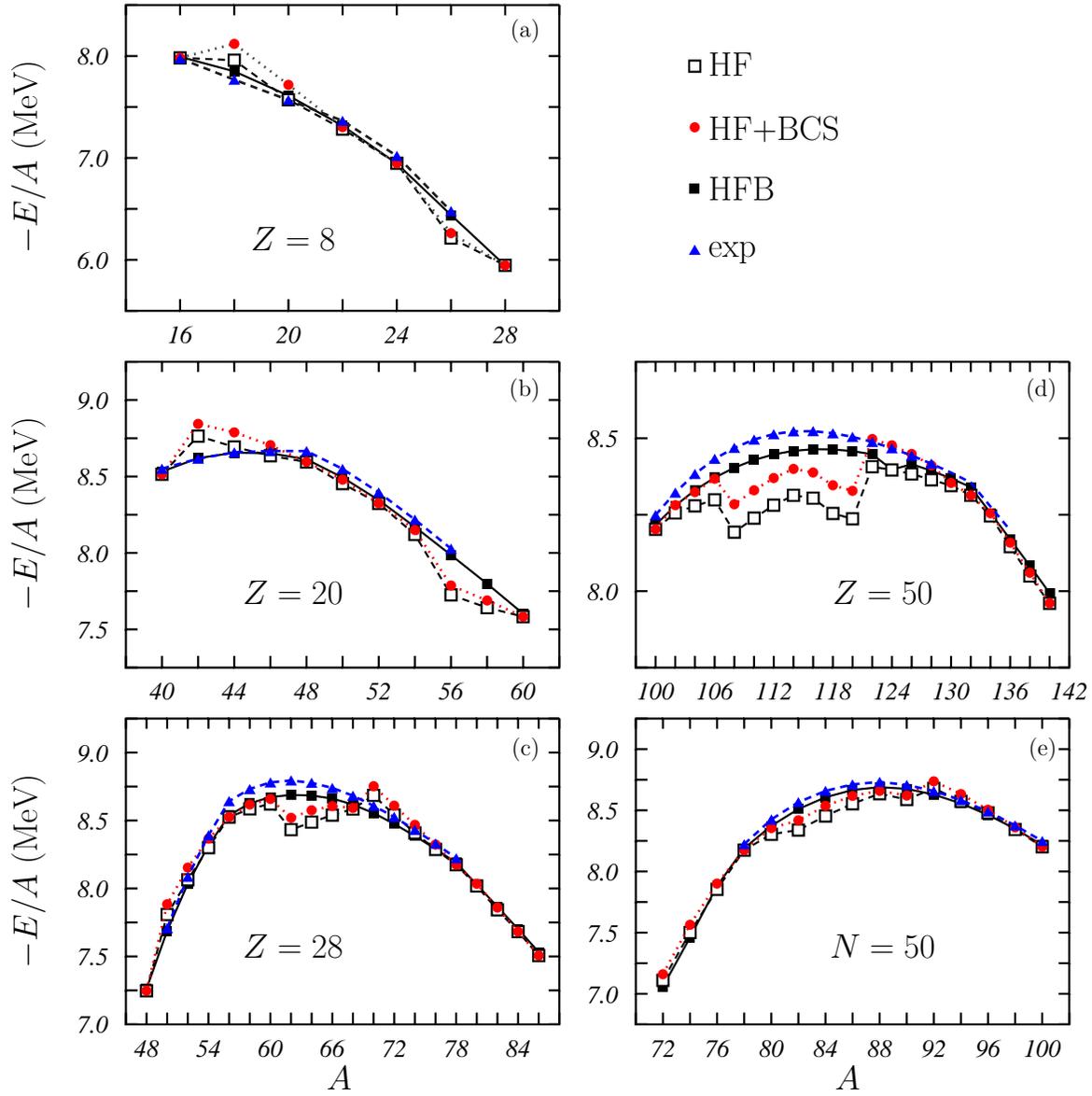} 
\caption{\small 
(Color on line) Binding energies per nucleon 
for oxygen (a),  calcium (b),  nickel (c), and tin (d) isotopes and  
$N=50$ isotones (e) calculated with the Gogny D1M in
HF+BCS (red circles), HFB (black squares) and HF (open squares)
models. The blue triangles show the experimental values taken 
from~\cite{aud03}. The lines have been drawn to guide the eyes.
\label{fig:EA}
}
\end{figure}

We show in figure \ref{fig:EA} the binding energies per nucleon for
oxygen, calcium, nickel and tin isotope chains and for a set of $N=50$
isotones. The red circles and the black squares indicate the HF+BCS and HFB results, respectively,
and the empty squares those of  the  HF calculation, where the pairing is not considered. 
The blue triangles show the experimental values taken from Ref.~\cite{aud03}. 
In this figure all the scales of the y axes have been chosen to
emphasize the differences between the results of the various calculations whose 
maximum relative value reaches 2\%. 

In all the panels we observe that the behaviour of the HFB results is quite smooth, 
while that of the HF+BCS calculations is less regular. 
In general, the HF+BCS results are closer to the experimental values than HF results except for $^{18}$O, 
$^{20}$O, $^{42}$Ca, $^{44}$Ca,$^{50}$Ni, $^{70}$Ni, $^{72}$Ni and $^{92}$Mo. Furthermore, HFB 
calculations are in better agreement with the experiment than our HF+BCS calculations that show again the
effects of a smaller pairing.

By considering in more detail the results shown in figure \ref{fig:EA},  we observe that our HF+BCS 
nickel binding energies, shown in the panel (c) 
have two discontinuity points related to the $^{62}$Ni and $^{70}$Ni nuclei. 
The discontinuities in our HF+BCS results are even more evident in the panel (d) where the binding
energies of the tin isotopes are presented. In this case,
the HFB and the HF+BCS values are in excellent agreement up to
$A=106$. After that, all the HF+BCS binding energies of the isotopes
going from $^{108}$Sn up to $^{120}$Sn are smaller than those of
the HFB calculations. We observe another discontinuity in coincidence with the $^{122}$Sn
isotope, where HF+BCS binding energy is larger than that 
obtained in the HFB calculation.
This trend continues up to the $^{126}$Sn nucleus, and, then,
the agreement between the results of the two calculations becomes
again excellent. 

\begin{figure}
\centering
\includegraphics[scale=0.6]{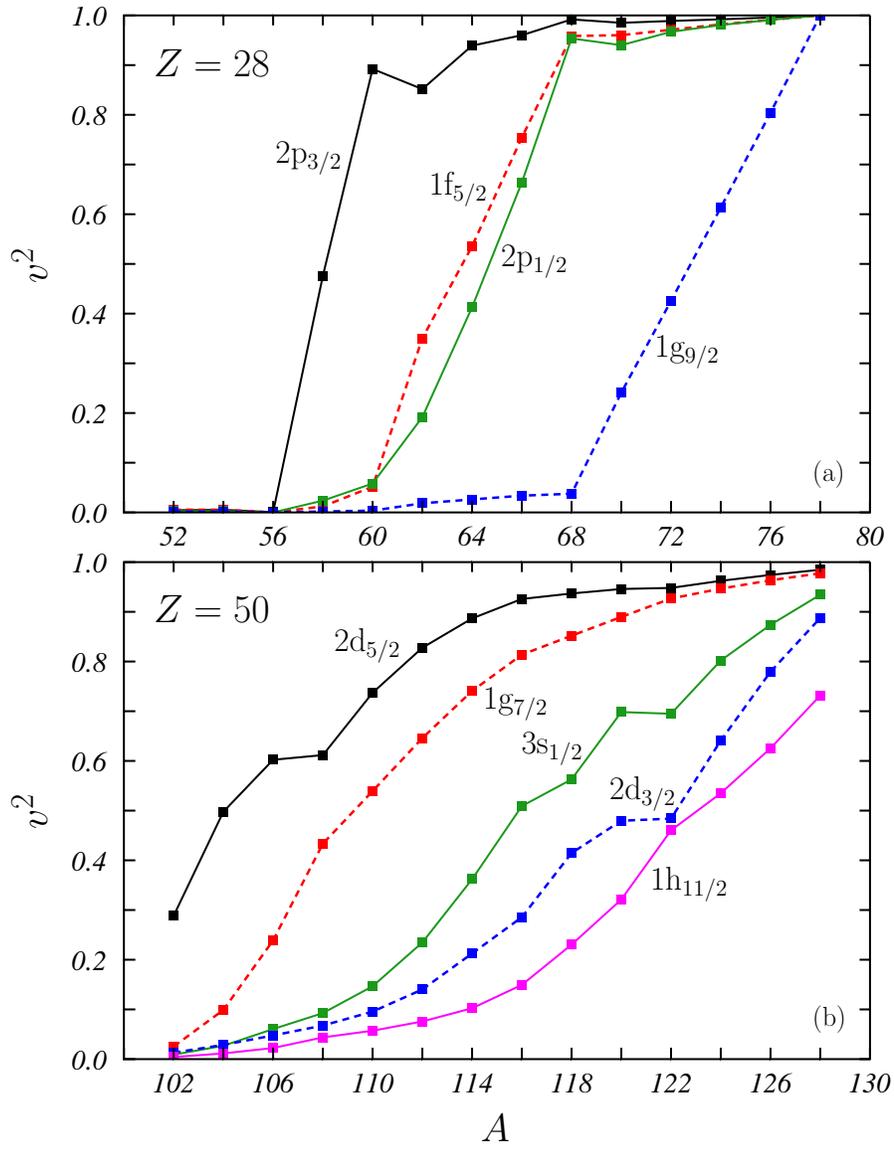} 
\caption{\small (Color on line) 
Occupation probabilities $v^2$ of different neutron s. p.
levels for nickel (a) and tin (b) isotopes obtained in HF+BCS
calculations with the D1M Gogny interaction.
\label{fig:v2tin}
}
\end{figure}

We have investigated the reason of this behaviour. After having
excluded any numerical instability of our calculations, we
focused our attention to the occupation of the
various s.p. levels. As example, we show in figure \ref{fig:v2tin} the
HF+BCS occupation probabilities $v^2$ for various neutron s.p. levels
of the nickel and tin isotopes around the Fermi level. In the nickel isotope chain, panel
(a), we observe a change in the behaviour of the occupation probability
for the 2p$_{3/2}$ state at $A=60$ and for the 1f$_{5/2}$ and 2p$_{1/2}$ states at $A=68$. 
In the case of the tin isotopes, panel (b), we also observe some
discontinuities in the trend of the $v^2$ values. This happens at $A=106$ for the
2d$_{5/2}$ state, and at $A=120$ for the 3s$_{1/2}$ and 2d$_{3/2}$
states. In these cases, the values of $v^2$ remain almost constant
for the next isotope and, after that, they start
again to rise. 

These behaviours
of the occupation probabilities $v^2$ are related to the energy of the s.p. HF 
levels. For example, from $^{58}$Ni to $^{64}$Ni,  
the energy of the neutron 2p$_{3/2}$ and 2p$_{1/2}$ states change by $-200$ 
keV at most. However, that of the 1f$_{5/2}$ state is reduced by more than 
600 keV from $^{60}$Ni to $^{62}$Ni, becoming almost constant afterwards. 
This provokes an enhancement of the occupation probability of the 
1f$_{5/2}$ state and, as a consequence, that of the 2p$_{3/2}$ is reduced. 
It is precisely between these two nuclei where the larger discrepancy between 
HF+BCS and HFB binding energies is observed. Similar arguments apply in the 
other cases above mentioned. This allows us to understand the discontinuities
pointed out in the \dndp \/ results of the panel (c) of figure \ref{fig:DN2}  
and of those we have observed in the panels (c) and (d) of figure \ref{fig:EA}.
These results indicate that our HF+BCS approach is less flexible in handling the filling of
the s.p. levels than the HFB method. 

The comparison of our HF+BCS results  with the experimental data, taken from the
compilation of Ref. \cite{aud03}, and indicated with blue triangles in
figure \ref{fig:EA}, is quite satisfactory.
In panel (a),   
we observe some differences with the experimental data in $^{18}$O
and $^{20}$O  that are both overbound, and in $^{26}$O which is, on the opposite, underbound. 
For the calcium isotopes, panel (b), our HF+BCS results overbind $^{42}$Ca and $^{44}$Ca. 
In the nickel chain, panel (c), we find a small underbinding in the isotopes
between the $^{56}$Ni and $^{68}$Ni, while  $^{50}$Ni, $^{70}$Ni and $^{72}$Ni are overbound.
In the panel (d) we observe that all the tin isotopes with $A<122$ are underbound, 
while a good agreement with the experimental data is found for the heavier isotopes. 
We observe an even better agreement in the $N=50$ isotone chain shown in panel (e). 
The maximum relative difference we found between the experimental data and our HF+BCS results 
is of about the 3\%.

\begin{figure}
\centering
\includegraphics[scale=0.8]{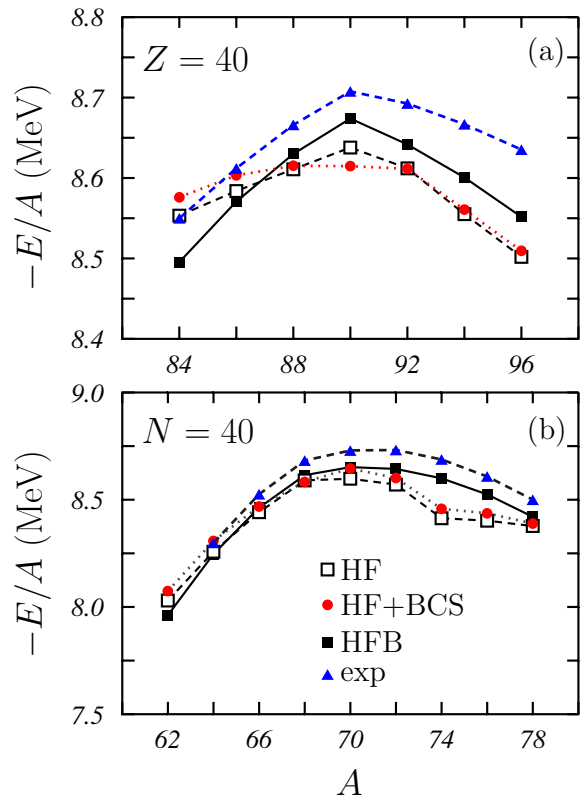} 
\caption{\small (Color on line) Binding energies for $Z=40$ isotopes (a) and $N=40$ isotones (b).
The meaning of the symbols is the same as in figure~\ref{fig:EA}.
\label{fig:EA40}
}
\end{figure}

We have calculated also the binding energies for nuclei where both proton and neutron shells are open.
The results for the $Z=40$ isotope chain, and for the $N=40$ isotone chain are shown
in figure \ref{fig:EA40}.
In these chains the $^{90}$Zr and $^{68}$Ni nuclei have closed the neutron and proton shells, respectively.

The results shown in this figure confirm the findings and the
observations we presented in the study of figure \ref{fig:EA}.
We find good agreement between HF+BCS and HFB results. 
The maximum relative difference is 1.7\%, found in the case of the
$^{74}$Se nucleus. Also the agreement with the experimental data is quite good.
In this case, the largest discrepancy we observe
is of about 2\%, always in $^{74}$Se. The effects of the pairing are 
rather small. The differences between HF+BCS and HF results are below 0.5\%.

\subsubsection{Radii.}
\label{sec:radii}

\begin{figure}
\centering
\includegraphics[scale=0.8]{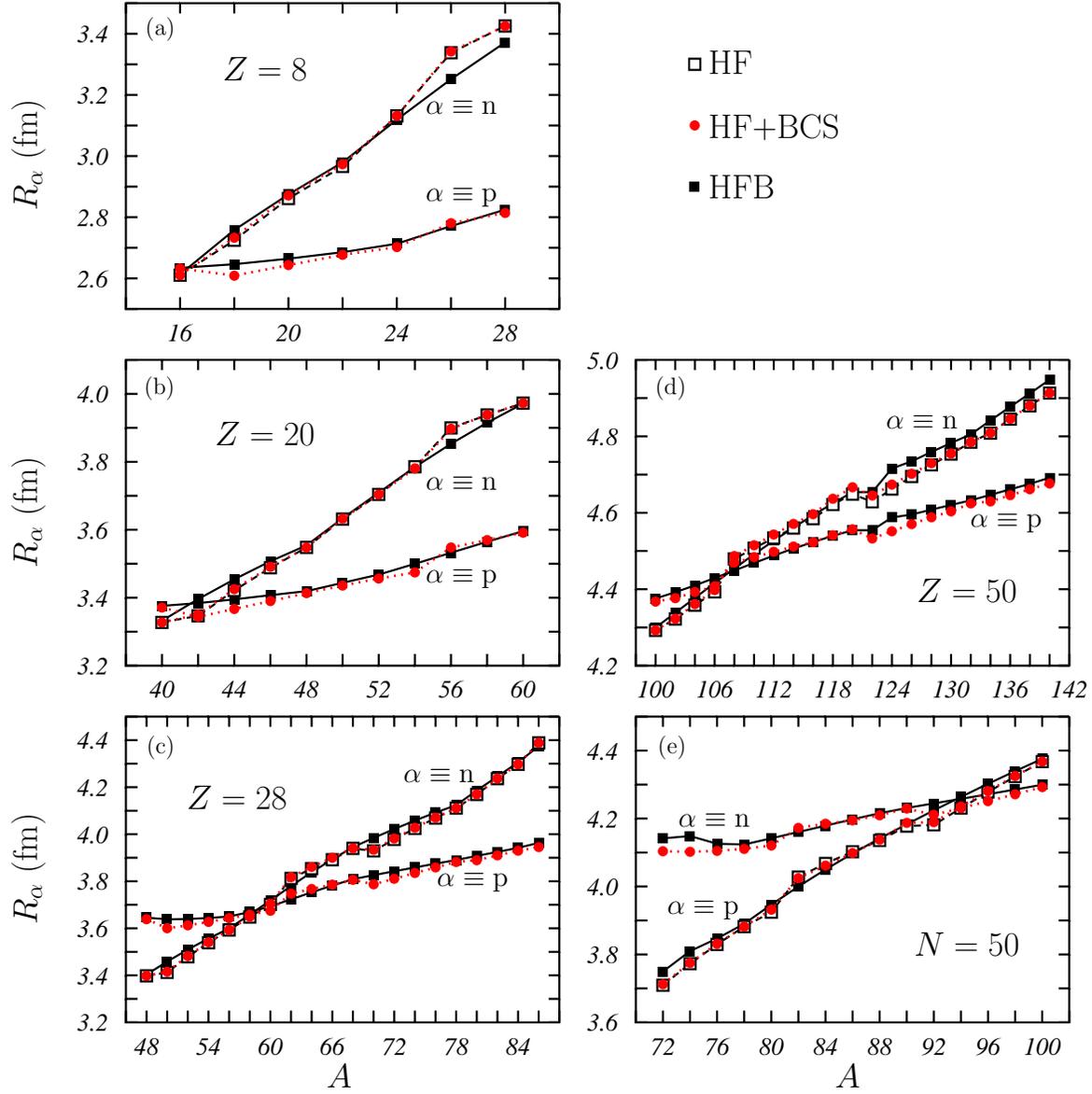} 
\caption{\small  (Color on line) 
Proton, $R_{\rm p}$, and neutron, $R_{\rm n}$, rms radii
for  oxygen (a), calcium (b),   nickel (c), and tin (d) isotopes and 
$N=50$ isotones (e). The red circles indicate the HF+BCS results, the  
black and open squares those obtained in HFB and HF calculations,
respectively. All the calculations have been carried out with the Gogny D1M
interaction. 
\label{fig:Rad-Z}
}
\end{figure}

\begin{figure}
\centering
\includegraphics[scale=0.8]{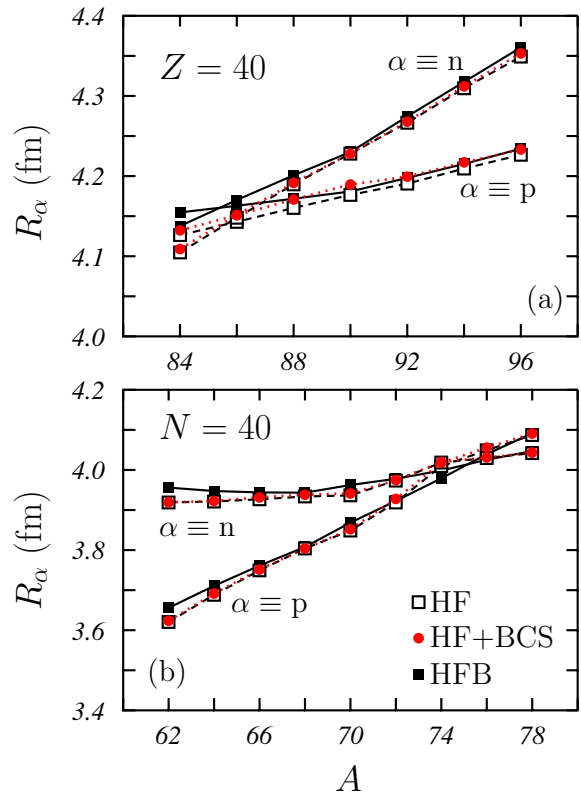} 
\caption{\small (Color on line) The same as in figure \ref{fig:Rad-Z}
  for $Z=40$ isotopes  (a) and  $N=40$ isotones (b).  
\label{fig:Rad-40}
}
\end{figure}

We show in Figs. \ref{fig:Rad-Z} and \ref{fig:Rad-40} 
the proton, $R_{\rm p}$, and neutron, $R_{\rm n}$, rms radii for
the various isotopes and isotones under investigation.
The red circles show the results
  obtained in HF+BCS calculations, the black and empty squares
  those obtained, respectively, by using the HFB and HF approaches.  
The results presented in Figs. \ref{fig:Rad-Z} and \ref{fig:Rad-40} have been obtained
with the D1M Gogny interaction.  
In the cases where the shells are closed, $Z=8$, 20, 28 and 50 for
protons and $N=50$ for the neutrons, HF and HF+BCS calculations 
give the same results, therefore we do not show
the empty squares of the HF results. The scales of the y axes have been chosen to emphasize the differences
between the various results.

We observe in Figs. \ref{fig:Rad-Z} and \ref{fig:Rad-40} 
a remarkable agreement between the HFB and HF+BCS results. 
We find a maximum relative difference of 1.5\%
for $R_{\rm p}$ in  $^{18}$O,  and of 1.9\%  for $R_{\rm n}$ in $^{26}$O. 
The comparison with the HF results indicates that the pairing effects
on the rms radii, are rather small.  The relative differences 
are smaller than 0.8\%.
 
\begin{figure}
\centering
\includegraphics[scale=0.8]{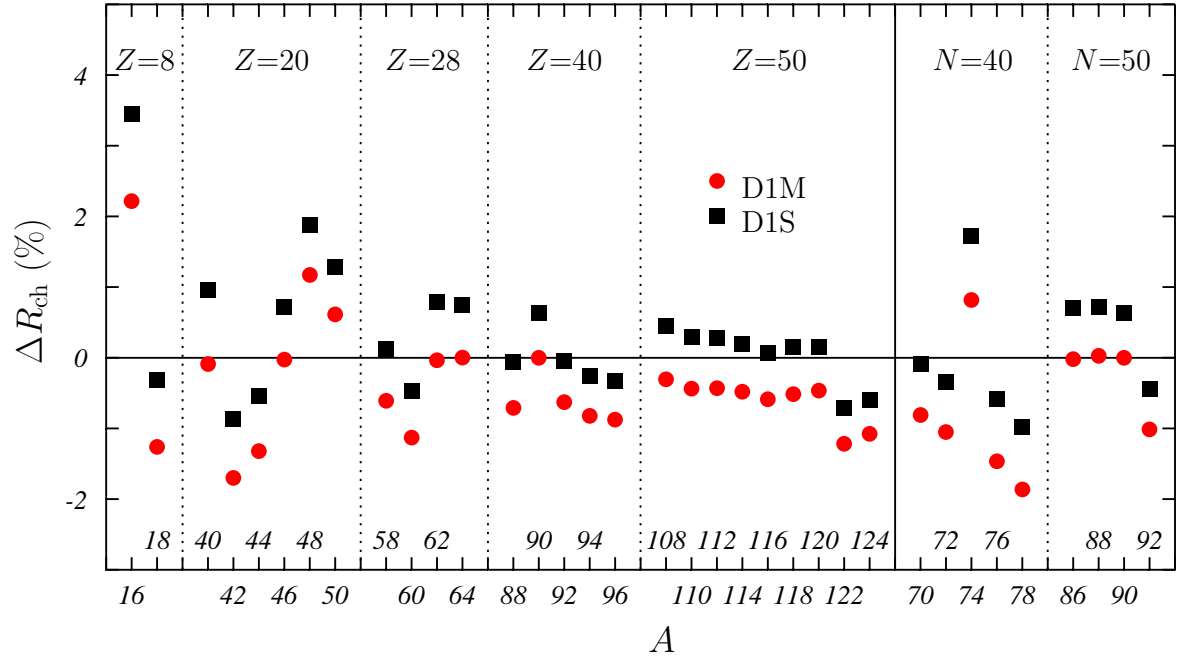} 
\caption{\small 
(Color on line) Relative differences, as given by
  equation (\ref{eq:charge-rad}), between HF+BCS results obtained with D1M (red points) and D1S (black squares)
  interactions and the experimental values taken from Ref. \cite{ang04a}.  
\label{fig:Rcharge}
}
\end{figure}

In order to compare our results with the experiment, we extract charge radii 
from the charge densities which we obtain 
by folding the corresponding proton densities with a dipole form factor.  
In figure \ref{fig:Rcharge} we show the relative differences, multiplied by 100, 
between the results of our HF+BCS calculations and 
the experimental radii taken from the compilation of Ref. \cite{ang04a}:
\begin{equation}
\Delta R_{\rm ch} (\%)\, = \, 100 \,
\, \displaystyle \frac{R^{\rm HF+BCS}_{\rm ch}-R^{\rm exp}_{\rm ch}}{R^{\rm exp}_{\rm ch}} \, .
\label{eq:charge-rad}
\end{equation}

In this figure we present the results only for those nuclei where experimental data  are available. 
We indicate with the red points the results obtained with the D1M
interaction, and with the black squares those obtained with the D1S force. 
All the relative differences are within the 2\%, with the only
exception of the $^{16}$O, where we found 2.2\% for the D1M
interaction and 3.4\% for the D1S force.
We notice that the radii obtained with the D1S force are always larger than those obtained with the
D1M interaction.

We expected pairing effects on the charge rms radii to be different from zero only
for the nuclei with $Z=40$ and $N=40$ and 50, where the proton s.p. levels in HF description are
not completely occupied. Even in these cases, they are rather small.
We observe relative differences between the radii calculated with HF+BCS and 
HF smaller than 0.3\%.

\subsubsection{Density distributions and electron scattering cross sections.}

We conclude our investigation by studying the effects of the pairing on the density
distributions of some selected nuclei.  The computational technique we use to solve the HFB equations \cite{ang00t}
does not provide us with wave functions, therefore we do not compare 
our results with those of HFB model.

The first case we discuss is that of two $N=20$ isotones, $^{36}$S and $^{34}$Si. 
In these nuclei the neutrons fill completely the s.p. states up to the 1d$_{3/2}$ level, 
therefore the pairing effects are present only in the proton sector, and 
we expect the proton density distributions to be sensitive to these effects.
The interest in these two nuclei has been raised by the fact that
some calculations \cite{nak12,gra09} indicate $^{34}$Si as a good candidate to be
a nucleus with bubble density profile, i. e. with a large depletion in its central part.
An experiment has been proposed by the collaboration GRETINA \cite{gretina}
at the Lawrence Berkeley National Laboratory, to produce the unstable 
$^{34}$Si nucleus and measure its proton density by means of
one-proton knock-out reaction in combination with a gamma
spectrometer  \cite{exp34Si}.

\begin{figure}
\centering
\includegraphics[scale=0.8]{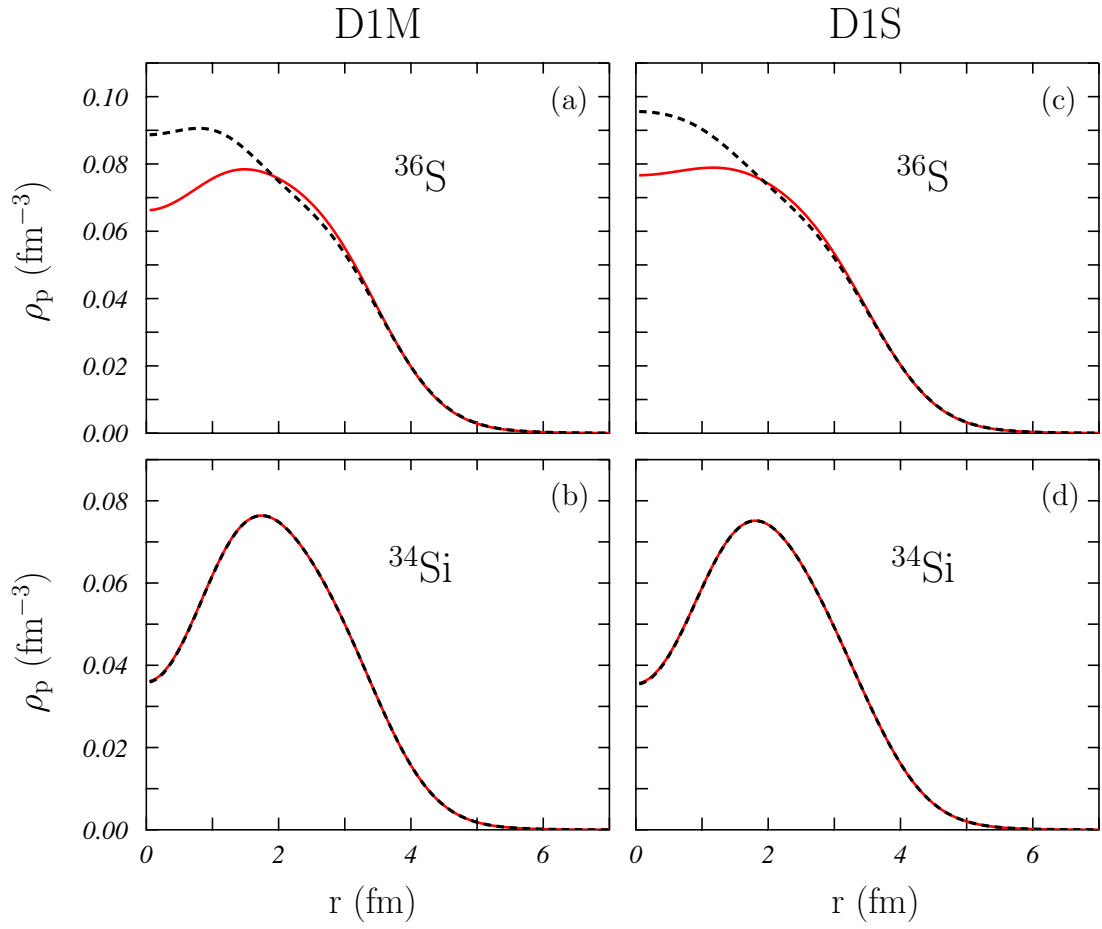} 
\caption{\small (Color on line) 
Proton densities distributions for $^{34}$Si and for $^{36}$S 
calculated in the HF (dashed black lines) and HF+BCS 
(solid red lines) approaches.  
The densities of the panels (a) and (b)
have been obtained with the Gogny D1M interaction, 
and those of the panels (c) and (d) with the D1S interaction. 
\label{fig:rho34Si36S}
}
\end{figure}

We present in figure \ref{fig:rho34Si36S} the proton densities of these two nuclei obtained by using the 
D1M and D1S interactions. 
The full red lines indicate the results obtained in HF+BCS calculations, 
while the dashed black  lines those of the HF calculations. 
Our results confirm the findings of Refs. \cite{nak12,gra09} 
and indicate that the proton density of $^{34}$Si has a bubble
structure.
In our density distributions the pairing effects are large in
$^{36}$S, but negligible in $^{34}$Si, independently of the interaction used in the calculation. 

These results are in agreement with those of
Ref. \cite{gra09} where a Skyrme SLy4 force has been used, together
with various pairing forces. The study of  Grasso et al. \cite{gra09}
concluded that the very large energy gap for $Z=14$, due to the
filling of the 1d$_{5/2}$ proton s.p. state, prevents pairing from
being relevant in $^{34}$Si, contrary to what happens in $^{36}$S. This occurs also in our HF+BCS calculations, where
the pairing effects strongly deplete the central part of the  $^{36}$S
proton distributions. In this nucleus we obtain 
a small bubble behaviour when the D1M
interaction is used. This detail disagrees with the results of  Nakada
et al. \cite{nak12} who found, with the D1M interaction, a behaviour
of the proton density more similar to the result we obtain with the D1S
interaction (see panel (c)).

\begin{figure}
\centering
\includegraphics[scale=0.7]{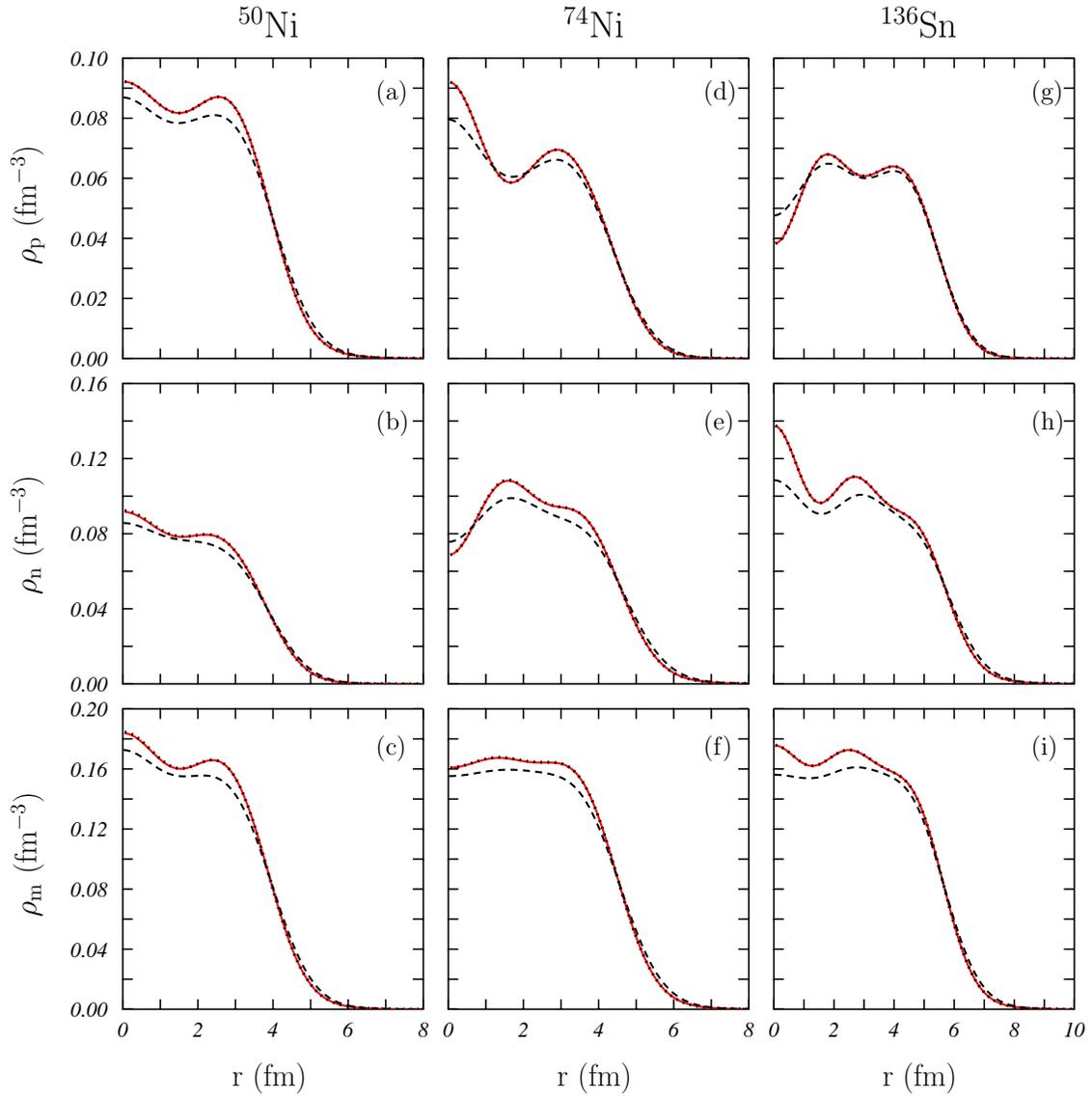} 
\caption{\small (Color on line) 
Proton, $\rho_{\rm p}$, neutron, $\rho_{\rm n}$  and matter, $\rho_{\rm  m}$, 
density distributions for the $^{50}$Ni, $^{74}$Ni and $^{136}$Sn 
nuclei obtained by using the Gogny D1M interaction in HF (dotted black
lines) and HF+BCS (solid red lines) calculations. We also show
the HF+BCS results (dashed black lines) of Ref.  \cite{sar07} obtained 
with a Skyrme Sly4 force. 
\label{fig:rhoNiSn}
}
\end{figure}

With the purpose of comparing our results with those of similar calculations 
done with different interactions, we have evaluated 
  the proton, neutron and matter density distributions of the
  $^{50}$Ni, $^{74}$Ni  and $^{136}$Sn nuclei. 
In figure \ref{fig:rhoNiSn} we compare our results, 
shown by full red lines, with those obtained in Ref. \cite{sar07}
by using the SLy4 Skyrme interaction. These latter results are 
indicated in the figure by dashed black lines. 
The agreement between our results and those
obtained with the Skyrme interaction is reasonably good
especially in the surface region.
In this figure, the dotted black lines indicate the densities obtained in HF calculations.  
In all the cases, these densities are almost coincident with the HF+BCS results.
In the nuclei we consider in the figure, pairing effects should appear in the
neutron and,  therefore, matter distributions. However, in these
cases, the open neutron shells, 1f$_{7/2}$ in $^{50}$Ni,
1g$_{9/2}$ in $^{74}$Ni  and 2f$_{7/2}$ in $^{136}$Sn, are rather
separated in energy (about 7, 5 and 3 MeV, respectively) with respect to the other open shells 
just above the Fermi surface. 
Because of this large energy separation the pairing effects on the densities are negligible. 
As a final remark on this figure, we observe that the large fluctuations in the
  nuclear interior shown by proton and neutron densities are smoothed
  out in the matter distributions which show a flatter behaviour.
  
As it is well known, the HF+BCS calculations may not be feasible for nuclei 
with a large neutron excess due to the appearance of an unphysical gas of 
neutrons around the nucleus. This spurious effect was observed by 
Dobaczewski \etal \cite{dob96} in HF+BCS calculations for $^{150}$Sn, 
performed with zero-range interactions. We have controlled that our 
results did not show this spurious effect.

\begin{figure}
\begin{center}
\includegraphics[scale=0.8]{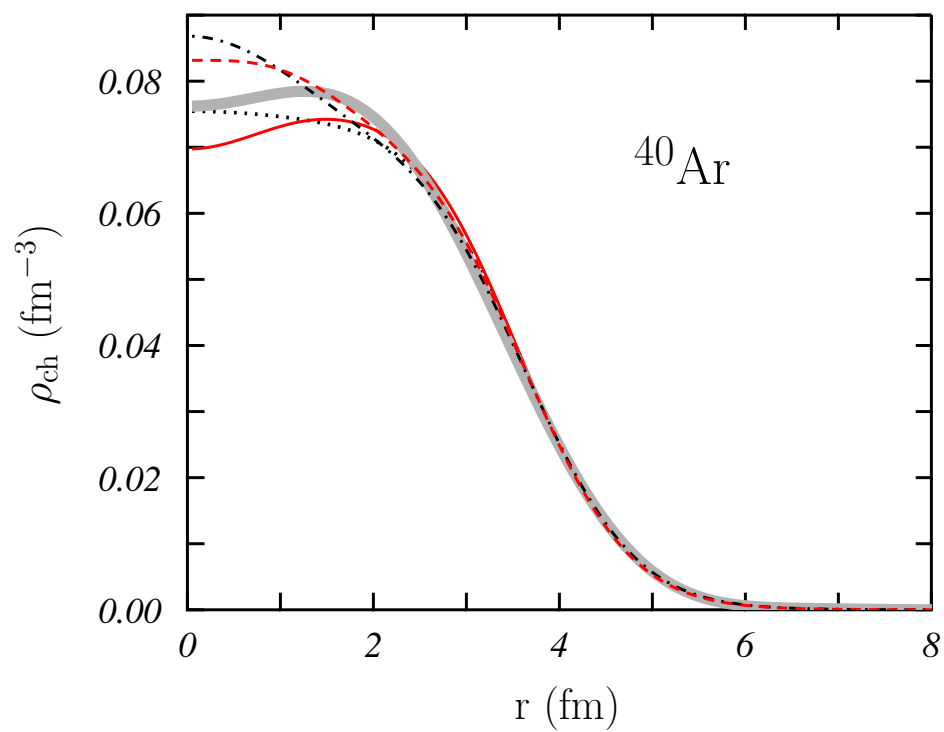} 
\caption{\small (Color on line) 
Charge density distribution of the nucleus $^{40}$Ar obtained in HF
with the D1M (dashed red line) and D1S (dashed-dotted black line) interactions
and HF+BCS with D1M (solid red line) and D1S (dotted black line) interactions,
compared with the experimental distribution  \cite{vri87}
(thick gray line).
}
\label{fig:rho40Ar}
\end{center}
\end{figure}

The nucleus $^{40}$Ar is quite interesting because
is the fundamental ingredient of liquid argon detectors used in the physics of 
the neutrinos and of rare events \cite{ame04,arn06,rub11}.
From the nuclear structure point of view the $^{40}$Ar is characterised by the fact that it 
has open shells in both proton and neutron sectors. The neutron shell that is open is the 1f$_{7/2}$
  level. As we have already pointed out for the case of the
   $^{50}$Ni nucleus, the energy difference between this state and the
  neighbouring ones is so large that pairing effects result to be
  negligible. For this reason we expect the pairing effects to be
  relevant only on the proton distribution.
In figure \ref{fig:rho40Ar} we show the charge densities obtained in 
HF calculations with D1M (dashed red line) and 
D1S (dashed-dotted black line) 
interactions and those obtained with the HF+BCS approach with D1M (solid red line) 
and D1S (dotted black line) interactions. 
The thick grey line shows the empirical charge density distribution \cite{vri87}. 
All the calculations reproduce rather well the 
  empirical distribution at the surface. The differences appear in the
  nuclear centre where the HF results rise continuously 
  while the empirical distribution shows a slightly decreasing behaviour. 
The pairing modifies the HF densities in the right direction,
producing, in the case of the D1M interaction, the depletion at the
centre shown by the experiment. 

\begin{figure}
\centering
\includegraphics[scale=0.7]{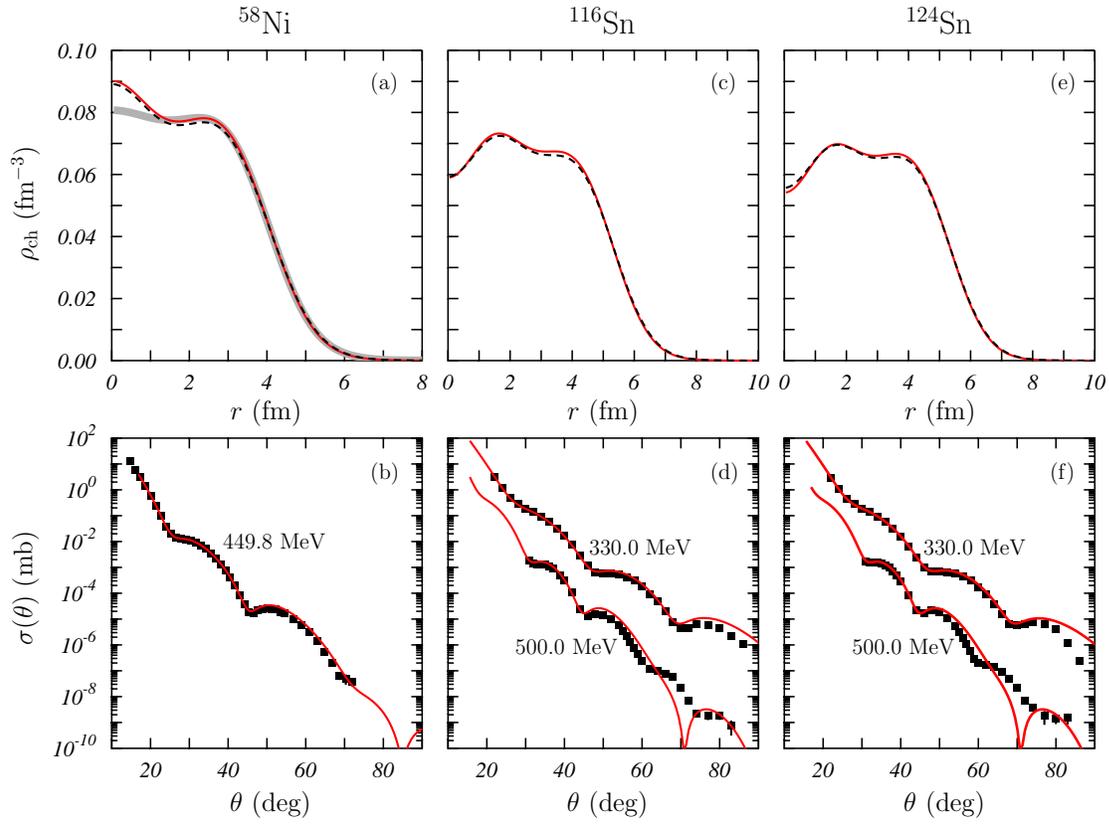} 
\caption{\small 
  (Color on line) Upper panels: Charge densities for $^{58}$Ni,
  $^{116}$Sn and $^{124}$Sn nuclei obtained in HF+BCS calculations  
  by using  the D1M (solid red lines) and D1S (dashed black lines) interactions.
  The thick gray line in panel (a) is the empirical charge density
  \cite{vri87}. Lower panels: Elastic electron scattering cross
  sections for the same nuclei calculated with the HF+BCS charge
  densities obtained by using the D1M interaction and compared with
  the experimental results \cite{fic70,fic72,cav80}. In each panel
  the electron energies are indicated. 
\label{fig:eeNiSn}
}
\end{figure}

To test the validity of our approach in the description of the  
density distributions, we use them to calculate
elastic electron scattering cross sections in
  distorted wave Born approximation \cite{ann95a,ann95b}.
We compare them with experimental data which are available  
for some of the open shell nuclei we have studied so far.
In the upper panels of figure~\ref{fig:eeNiSn} we show the charge densities obtained with
the D1M and D1S interactions, indicated, respectively, by the full red  and dashed black lines,
for $^{58}$Ni, $^{116}$Sn and $^{124}$Sn. In the panel (a) of the figure we show with a thick gray line
  the empirical charge density \cite{vri87}.

In these nuclei the proton levels are completely filled, therefore the pairing effects 
on the charge distributions are very small. The HF density distributions almost coincide with those 
obtained in HF+BCS calculations. Also the differences between the results of the 
calculations done with different interactions are 
rather small, as we show in the upper panels of figure~\ref{fig:eeNiSn}.

The elastic electron scattering cross sections shown in the lower
panels of the figure have been calculated by using the charge
densities obtained with the D1M interaction. Our results describe
rather well the experimental data of Refs. \cite{fic70,fic72,cav80}
for the three nuclei considered. The agreement is particularly good
for the $^{58}$Ni nucleus, panel (b).

For $^{116}$Sn and $^{124}$Sn nuclei, the agreement is very good for
the data taken at 330.0 MeV, and it is slightly worse for those taken
at 500.0 MeV for scattering angles larger than 50 degrees. The
increasing resolution power of the probe emphasises the differences
with the experimental charge distributions, differences which we
expect to show up mainly in the nuclear interior
\cite{ann95a,ann95b}.

\section{Summary and conclusions}
\label{sec:conclusion}

We have developed a fully self-consistent HF+BCS approach where the same finite-range interaction
is used in the two steps of the process, first the HF calculation generating the set of s.p. wave functions
and energy, and, second,  the BCS calculation taking care of the pairing correlations between these 
s.p. states. Since the interaction has a finite-range, we obtain a natural convergence of the BCS results
once the dimension of the s.p. configuration space have reached a certain size. 
In our calculations we use two interactions 
of Gogny type, the D1M and D1S forces.  

The motivation of our work is to construct a parameter free computational scheme which 
can be applied to investigate all the regions of the nuclear chart, therefore, with a large prediction power. 
Our HF+BCS approach is competitive with the HFB model, which is more elaborated from both
the theoretical and computational points of view. One of the main purposes of this work was to test
the validity of our HF+BCS model by comparing our results against those of the HFB calculations.
This study has been carried out mainly by investigating 
the particle number fluctuation indexes \dnd, and we found an overall
agreement between the results of the two approaches. There are, however, specific cases showing
non-negligible differences, cases we have analysed in detail. We have also studied differences between 
HF+BCS and HFB results in other quantities such as binding energies and rms radii, and we found 
these differences to be even smaller than those found for the \dnd values. 
We can summarize our findings by saying
that  all the cases we have investigated show that pairing 
correlations are larger in HFB than in HF+BCS results. This indicates a certain rigidity of 
the BCS calculations to mix the contribution of various s.p. levels, with respect to the HFB approach.

Our model, applied in various region of the nuclear chart for isotope and isotone chains from 
oxygen up to tin, indicates that pairing affects the total binding energies by few percent. We found a 
maximum relative difference of the 2.3\% in the nucleus $^{56}$Ca. The effects on the rms and 
charge radii are even smaller, well below the 1\%.
We find that the effects of pairing on the density distributions are mainly located in the nuclear interior. 
Our approach describes rather well the available empirical densities and the experimental elastic
electron scattering data.

In conclusion, we can state that the relative simplicity of our HF+BCS approach is paid in terms
of flexibility of the pairing correlations in mixing the contribution of various s.p. levels. This implies
an underestimation of the pairing effects with respect to the more elaborated HFB approach. 
Apart from some specific cases where the HF s.p. levels are fully occupied, from the quantitative point
of view, the differences between the results of the two approaches are not so large to destroy the 
validity of our HF+BCS model. Because of its relative simplicity, 
our model allows us to disentangle well the effects of the various components of the
calculations such as interaction, s.p. energies and wave functions, and pairing correlations.

The purpose of the present work was to analyze the capabilities of the simplest HF+BCS approach. 
Nevertheless, this approach could be improved by considering a procedure where the occupation 
numbers obtained after a first BCS calculation are used as input for a new HF calculation on 
top of which the BCS is applied again. In this way an iterative, self-consistent approach could 
be carried out and a better agreement with HFB results may be obtained.

\ack{This work has been partially supported by the PRIN
(Italy) {\sl  Struttura e dinamica dei nuclei fuori dalla valle di stabilit\`a},
by the Spanish Ministerio de Ciencia e Innovaci\'on under contract
FPA2012-31993 and by the Junta de Andaluc\'{\i}a (FQM0220).}


\pagestyle{empty}
\clearpage
\newpage
\newpage
\clearpage

\clearpage
\clearpage
\clearpage
\clearpage
\end{document}